\documentclass[aps,prd,twocolumn,noshowpacs,superscriptaddress,groupedaddress,nofootinbib,preprintnumbers]{revtex4}  
\pdfoutput=1
\usepackage[T1]{fontenc}
\usepackage[latin9]{inputenc}
\usepackage{graphicx}  
\usepackage{dcolumn}   
\usepackage{bm}            
\usepackage{amsmath,amssymb,amsfonts,wasysym} 
\usepackage{mathtools,empheq} 
\usepackage{color}
\usepackage[english]{babel}
\usepackage{bbm}
\usepackage[]{hyperref}%
\hypersetup{
colorlinks=true,
linkcolor=red,
anchorcolor=black,
citecolor=blue,
filecolor=cyan,
menucolor=red,
runcolor=filecolor,
urlcolor=magenta,
bookmarks=true,
bookmarksnumbered=true
}

\makeatletter
\def\l{\label}

\def\({\left(}
\def\){\right)}
\def\f{\frac}
\def\be{\begin{equation}}
\def\ee{\end{equation}}
\def\bry{\begin{array}}
\def\ery{\end{array}}
\def\bes{\begin{subequations}}
\def\ees{\end{subequations}}
\def\bit{\begin{itemize}}
\def\eit{\end{itemize}}
\def\ben{\begin{enumerate}}
\def\een{\end{enumerate}}
\def\dst{\displaystyle}

\def\ovl{\overline}

\newcommand{\Dsl}{D\llap{/\kern+1.5pt}}

\newcommand{\tev}{\textrm{ TeV}}
\newcommand{\gev}{\textrm{ GeV}}

\usepackage{slashed}
\usepackage[usenames,dvipsnames]{xcolor}
\renewcommand{\url}[1]{\href{#1}{\textsc{\textcolor{Emerald}{WebSite}}}}

\makeatother

\begin{document}

\preprint{DFPD-2012/TH/26}
\preprint{IFT-UAM/CSIC-12-118}
\preprint{LPN12-135}
\preprint{UMD-PP-012-025}

\title{RPV stops bump off the background}

\author{Roberto Franceschini}
\email{rfrances@umd.edu}
\address{Department of Physics, University of Maryland, College Park, MD 20742-4111, USA}
\author{Riccardo Torre}
\email{riccardo.torre@pd.infn.it}
\address{Instituto de F\'isica Te\'orica UAM/CSIC,
Universidad Aut\'onoma de Madrid, Cantoblanco, E-28049, Spain}
\address{Dipartimento di Fisica e Astronomia, Universit\'a di Padova and INFN Sezione di Padova, Via Marzolo 8, I-35131 Padova, Italy}
\address{SISSA, Via Bonomea 265, I-34136 Trieste, Italy}

\begin{abstract}
We study the 8~TeV LHC reach on pair produced heavy flavored di-jet resonances. Motivated by theories of R-parity violation in supersymmetry we concentrate on a final state with two $b$-jets and \mbox{two light} jets. 
We exploit b-tagging to reject the background and discuss its importance at the trigger level to probe light stops. 
We present kinematical selections that can be used to isolate the signal as a bump in the mass distribution of the candidate resonances.
We find that stops with R-parity violating couplings giving rise to fully hadronic final states can be observed in the current run of the LHC.
Remarkably, the LHC can probe stop masses well within the range predicted by naturalness. 
\end{abstract} 

\maketitle

\section{Introduction}

A new boson has recently been discovered at the LHC \mbox{\cite{The-CMS-Collaboration:2012ys,The-ATLAS-Collaboration:2012rt}}.
From its production cross section and a preliminary assessment of
its properties it is favored to be significantly coupled to the SM
matter in a way similar to the SM Higgs boson.

The existence of such a SM-like boson adds new emphasis on a long standing
question about the existence of Supersymmetry (SUSY) at the TeV scale. In fact, at
this stage there are indications that, if any, such TeV scale SUSY
is likely to be non-minimal. An indicator of the non-minimality of
the realization of SUSY at the TeV scale is the fact that the mass
of the new boson is too large to be  \emph{naturally} accommodated in
the Minimal Supersymmetric Standard Model (MSSM) and even in the Next-to-MSSM (NMSSM)~\cite{Hall:2011rt,Agashe:2012eu,Draper:2011aa}.
We take this input as a further motivation for non-minimal scenarios
of supersymmetry.

Departing from the minimal set-up, we think that one should still insist on naturalness 
 in formulating SUSY models that can be probed at the LHC.
Naturalness requires that the
states with strong interactions with the Higgs sector must be light,
hence they are the minimal degrees of freedom of the Effective Natural
SUSY theory relevant at LHC energies. The minimal set of states that
are expected to be light are the stop and the sbottom squarks, the
higgsinos and the gauginos~\cite{Dimopoulos:1995if,Cohen:1996vc}.

The first and second generation squarks and the sleptons, on the other
hand, do not need to be light by naturalness arguments. This is particularly
relevant considering the strong bounds of the LHC on SUSY spectra
containing light squarks~\cite{CMSSusySummary,CMS-PAS-SUS-11-016}.

The results of the early LHC on ``benchmark'' SUSY spectra
and the great relevance of Natural SUSY from the theory perspective~\cite{Barbieri:2009kx,Brust:2011gf,Essig:2012ew,Papucci:2011rc,Kats:2012hb,Allanach:2012ai}
motivated a recent effort to search for third generation
squarks.
These searches are now constraining significant portions of the parameter
space of Natural SUSY and in many scenarios the mass of the stop squark
is bound to lie in a strip around $m_{t}$~\cite{ATLASSusyDirectStopSummary}.
More precisely a large part of the not yet excluded parameters space
lies close to the line \sloppy\mbox{$m_{\tilde{t}}\sim m_{\chi}+m_{t}$}, where
$m_{\chi}$ is the mass of the Lightest Supersymmetric Particle (LSP)
supposed to be stable and invisible to the detectors in all searches.

This latter assumption on the nature of the LSP is central in the
design of current analyses. In fact it defines the main limitation
of the current searches and the scope of several proposals, including
ours, to extend them.

In the context of models with a stable LSP, several ideas have been put
forward to extend the sensitivity of present experiments~\cite{Han:2012ve,Bai:2012ul,Kaplan:2012ly,Plehn:2012mz,Alves:2012gf,Yu:2012uq,Kilic:2012yq}.
However, despite the relatively light mass,
a stop with a mass around the one of the top quark, \emph{i.e.}~at the
heart of the naturalness region, remains challenging to observe\footnote{See also Ref.$\,$\cite{Kats:2011fk} for bounds and discovery potential of a meta-stable stop at colliders.}. The main difficulty to overcome is typically that the final states are too ``top-like''  and therefore difficult to be distinguished from the usual SM sources. Particularly
problematic is the fact that the new physics events lack large missing
energy in the final state, which is one of the most effective selections
of the current SUSY searches.

The difficulty to exclude new physics, even with substantial cross section,
when it does not give rise to large missing energy indicates the status of where ongoing
searches for SUSY and Natural SUSY currently stand.

\bigskip

In this work we discuss the discovery potential of scenarios
where the large missing energy is completely absent due to no additional assumptions on the stability of the LSP. We consider decays of the LSP mediated by coupling that violate R-parity~\cite{1982PhLB..119..136A,Hall:1983id,Ross:1984yg,PhysRevD.40.2987},
which must generically be small not to be in contrast with observations
(for a review on R-parity breaking and bounds on the various couplings
see Refs.~\cite{Barbier:2005rr,Dreiner:1998iz,Bhattacharyya:1998ma,Kao:2009ve}). 

Despite the R-Parity Violating (RPV) couplings being small they can significantly alter
the phenomenology of the lightest supersymmetric particle\footnote{See, for instance, Ref.~\cite{Dreiner:2012qy} for a recent study of
the decay patterns of supersymmetric particles in presence of RPV.%
}.
We  focus on the case where the LSP is a stop squark that decays
into a $b$ and an $s$ quark
\be\l{decay}
\tilde{t}\to bs\,.
\ee
%
We consider small RPV couplings to easily satisfy flavor constraints.
This also allows us to neglect the single-resonance production of superpartners
and to rely only on the model-independent pair-production from
color gauge interactions. We assume however that the RPV couplings are large enough to give  prompt decays of the stop\footnote{For results on the case of long lived sbottom and stop giving rise to mesino-antimesino oscillations see Refs.~\cite{Sarid:2000lr,Berger:2012yq,Evans:2012oq}.}.

We consider the flavor pattern of Eq.~\eqref{decay} because the RPV coupling
involving the $t$, $b$ and $s$ flavors is among the least constrained and because
it is motivated by theories that describe R-parity breaking in a predictive and controllable way~\cite{Nikolidakis:2008tu,Smith:2008no,Csaki:2012zr,Keren-Zur:2012cr,Ruderman:2012qy}.

The flavor content of the final state in the decay of Eq.~\eqref{decay} allows background
rejection using $b$-tagging techniques. Furthermore the fact that no new
particle in the process is undetected opens the possibility to reconstruct
the masses of the resonances and to improve the background rejection. 

We  consider stop masses above 100$\,$GeV as dictated by previous
searches of the RPV stop~\cite{OPALCollaboration:2003hu,CDFCollaboration:2013up,Beringer:1900zz}.
Besides the lightest stop we are not assuming other light superpartners,
therefore our result fills the gap left by other recent studies that
assumed a light sbottom-stop doublet at the bottom of the SUSY spectrum.
In fact in our study we do not exploit the leptons coming from
electroweak currents that can dramatically help to reject the SM background.
Hence our study applies to the two cases where either one or both the
SU(2) singlet and doublet stop are light, including the case of a
compressed third generation squark spectrum with $m_{\tilde{b}_{1}}\gtrsim m_{\tilde{t}_{1}}$
or $m_{\tilde{b}_{1}}\simeq m_{\tilde{t}_{2}}\gtrsim m_{\tilde{t}_{1}}$.

The LHC experiments have already performed searches for scenarios where
new physics does not give rise to large missing energy. For instance, they
have been looking for the pair-production of colored resonances in the multi-jet final states~\cite{ATLAS-Collaboration:2012kc,Aad:2011ly,CMS-PAS-EXO-11-016,CMSCollaboration:2013ui,CMS-PAS-EXO-11-001,CMS-PAS-EXO-11-075,The-CMS-Collaboration:2012fi}\footnote{Further extensions of these searches to exploit the production regime
of boosted resonances have also been proposed~\cite{Curtin:2012cr,Han:2012ys}.%
}. In particular, in Refs. \mbox{\cite{Aad:2011ly,CMS-PAS-EXO-11-016,ATLAS-Collaboration:2012kc,CMSCollaboration:2013ui}},
searches have been presented for a final state with 4 hard jets, very similar to the one
that we expect from 
\be\l{process}
pp\to\tilde{t}\ovl{\tilde{t}}\to b\bar{b}s\bar{s}\,. 
\ee
The current analyses, however, are not sensitive to the presence of
heavy flavors in the final state and do not place any bound on
a triplet of color $SU(3)$ such as the stop decaying into light jets~\cite{Evans:2012oq}.

Other studies of similar new physics objects appeared in Refs.~\cite{Schumann:2011fj,Bai:2011nx,Kilic:2009wd,Del-Nobile:2010oq,Plehn:2009yq,Choi:2009kx}.
However in these works no flavor structure of the final
state could be exploited to isolate signals. As such, our case of a flavored
di-jet resonance from the stop is crucially different. Furthermore
we develop a selection strategy significantly different from
those discussed in earlier works (mainly in the identification of
the resonance candidates, see Section~\ref{search strategy}).

A more directly relevant earlier work is the one of Ref.~\cite{Choudhury:2011qt}
where hadronic RPV stop decays have been considered at the 14 TeV
LHC. However the range of masses of interest for that work is rather
shifted towards a higher and unnatural region. Furthermore we believe
that a discussion of the sharpness of the isolation of the searched
di-jet peak is in order for a fully hadronic final state such as the
one produced in the process of Eq.~\eqref{process}. In fact we  present the analysis
putting the emphasis on the shapes of the background and the signal distributions.
Finally, in contrast to Ref.~\cite{Choudhury:2011qt}, we  concentrate on the discovery
chances of the ongoing run of the LHC at 8 TeV, therefore considering an integrated luminosity of around $20$~fb$^{-1}$.


Compared to previous theoretical works we can use the experience  from recent ATLAS~\cite{Aad:2011ly,ATLAS-Collaboration:2012kc} and CMS~\cite{CMS-PAS-EXO-11-016,CMSCollaboration:2013ui} searches to compare two different approaches for the candidate resonance reconstruction. We also use experience from data to address the issue of triggering
on interesting events from ``light'' particles such as a stop of a few
hundred GeV in the extremely complex and noisy environment of the
LHC. In particular we  discuss how \mbox{$b$-tagging} at trigger
level, \emph{$b$-trigger} in the following, can be used to have sustainable
trigger rates while searching for low mass resonances, such as the light
stop predicted by Natural SUSY.

With our work we intend to motivate experimental activity to search
for the RPV stop, that is an under-investigated, though well motivated,
case of a generic top partner, \emph{i.e.}~a partner of the top quark that, to
stabilize the electroweak scale, is expected to have mass and couplings to
the Higgs boson similar to those of the top quark. In this sense our
work goes in the same direction of the recent work of Ref.~\cite{Brust:2012yq}
where direct pair-production of sbottom squarks or heavy stops decaying
into lighter stops $\tilde{b}\to\tilde{t}_{1}W^{(*)}$, $\tilde{t}_{2}\to\tilde{t}_{1}h^{(*)}$,
$\tilde{t}_{2}\to\tilde{t}_{1}Z^{(*)}$, where $\tilde{t}_{1}\to jj$,
has been studied in the context of RPV SUSY\footnote{See also Ref.~\cite{Graham:2012qe} for a study of the phenomenology of
RPV SUSY with long lived LSP decaying at a measurable displaced distance
from the primary interaction point.%
}. 

We remark that our search strategy is mostly based on general
considerations on the pair-production of di-jet resonances and therefore,
except for differences in the selection efficiencies, is expected to also
apply to other kinds of heavy-flavored di-jet resonances. For instance
the analysis could be applied to heavy fermionic partners of the bottom
quark when they decay mostly as $B\to bg$  via a dimension 5 chromo-magnetic
interaction. Hence, although being inspired by the search for the RPV
stop, we can anticipate that the proposed experimental search can
have a much wider reach.

\bigskip

Our paper is organized as follows. In Section~\ref{sec:R-parity-breaking}
we  review the motivations for R-parity and we remind how
these motivations are weakened once Natural SUSY is taken as an effective
theory up to relatively low scales. We  also briefly review possible
ways to break R-parity in a controlled way. In Section~\ref{search strategy} we describe our search strategy
for the RPV stop and estimate the results of these searches at the LHC
with $\sqrt{s}=8\tev$ and an integrated luminosity of 20 fb$^{-1}$. In Section~\ref{conclusions} we conclude.

\section{R-parity and its breaking\label{sec:R-parity-breaking}}

R-parity is a symmetry originally introduced to overcome the issues
with baryon and lepton number conservation that arise when the SM
is extended to be supersymmetric. With the minimal
superfield content needed to make the SM supersymmetric~\cite{Dimopoulos:1981zb},
the model contains baryon number, $B$, and lepton number, $L$, violating terms
already at the renormalizable level.
These are the super-potential terms
\begin{equation}
W_{\slashed B}=\lambda''UDD\,,\label{eq:Wbaryon}
\end{equation}
and
\begin{equation}
W_{\slashed L}=\lambda LLE+\lambda'QLD+\mu'LH_{d}\,.\label{eq:Wlepton}
\end{equation}
From a theory perspective
it looks appealing to forbid these terms by some dynamics, without explicitly
imposing baryon and lepton number conservation, as it happens
in the SM, where baryon and lepton number conservation arise as accidental symmetries from the gauge and Lorentz structure of the theory and its field content
(in perturbation theory and barring small effects from possible Majorana
masses for neutrinos). 

Requiring conservation of R-parity, defined as \sloppy\mbox{$R_{P}=(-1)^{2S+3(B-L)}$}, where $S$ denotes the spin,
the renormalizable superpotential terms of Eqs.~\eqref{eq:Wbaryon} and \eqref{eq:Wlepton}
are forbidden and the effects of the violation of baryon and lepton
numbers are pushed to the level of higher dimensional operators~\cite{Sakai:1981pk,PhysRevD.26.287}
of the type
\[
W_{\text{HDO}}\propto UUDE\,.
\]
In SUSY models where all the mass scales beyond the TeV
scale are close to the GUT scale R-parity is a viable symmetry
to forbid too large baryon and lepton number violation. 

In the context of effective theories valid up to some intermediate
scale well below the GUT scale the motivation for R-parity becomes
weaker. In fact the higher dimensional operators, which in general are
allowed by R-parity and mediate $\slashed B$ and $\slashed L$
transitions, may be suppressed by a too low scale to explain the stability
of the proton. In this sense in a low scale SUSY model \mbox{R-parity} is not
sufficient to make the model phenomenologically viable, and hence not
a necessary ingredient of the model~\cite{Brust:2011gf}. Other mechanisms need to be invoked to suppress the large $\slashed B$ and $\slashed L$ transitions.

Given the rather strong restrictions on the phenomenology implied
by R-parity, we think that in the context of an effective theory for
Natural SUSY, valid only up to 10-100 TeV, it
is very well motivated to remove the assumption of R-parity conservation.

Remarkably the breaking of R-parity can be motivated also from a theory
standpoint when the stringent requirements of flavor physics experiments
are taken as a guideline for SUSY model building both in the context
of elementary and composite Higgs models \mbox{\cite{Nikolidakis:2008tu,Smith:2008no,Csaki:2012zr,Keren-Zur:2012cr}\footnote{
A collective breaking of R-parity has also recently been studied in Ref.~\cite{Ruderman:2012qy}.}.} 
In these constructions the large mass of the top quark singles it
out from the other lighter quarks.
As a consequence of its special nature, and happily matching the available
constraints from experiments, the third generation can host larger
RPV couplings. This
is for instance the case in the decay of the $\tilde{t}$ which, in
the mentioned scenarios, has a preference to decay into heavy quarks.

\section{Search Strategy \label{search strategy}}
\subsection{$b$-tagging and triggers}
A crucial issue to search for new physics in multi-jet final states is that of triggering.
Due to the absence of other hard objects in the event the new physics
must pass multi-jet triggers or ``inelasticity'' triggers such as a trigger on the scalar sum of the jets $p_{T}$.
 
The large cross section of QCD multi-jet production and the high instantaneous luminosity of the LHC forces us to put rather high thresholds for these fully hadronic triggers. This is illustrated well, for instance, by the evolution of the ATLAS searches. In Ref.~\cite{Aad:2011ly} the ATLAS Collaboration used low luminosity 2010 data and a first level trigger requiring at least four jets with a transverse momentum, $p_{T}$, of at least 5 GeV. With that trigger new physics searches were carried out with $p_{T}$ of the jets as low as 55 GeV. Instead in the search of Ref.~\cite{ATLAS-Collaboration:2012kc} they used the much larger 2011 dataset and had to use  a multi-jet trigger requiring at least four jets that had an almost flat efficiency for $p_{T}\geq 80\gev$. 
 
The need to raise the trigger thresholds makes it difficult to search for light states, that are less likely to give jets with sufficiently high transverse momentum. In fact the scale of the $p_{T}$ of the jets  from the decay of the pair produced new physics resonances is typically a fraction of the mass of the resonance. 

Hereafter we  consider the possibility of relying on the peculiarity of $b$-quarks to deal with fully hadronic final states in a high luminosity regime, still keeping relatively low $p_{T}$ thresholds for the trigger. The possibility to use raw $b$-tagging information at the trigger level in a high luminosity environment has already been demonstrated in several measurements of properties of the Standard Model, as for instance in the study of hadronic decays of the top quark~\cite{ATLAS-CONF-2012-032,ATLAS-CONF-2012-031}. Thus, in the following we  assume that the data has been recorded using a multi-jet trigger with one or more $b$-tags at trigger level. However it should be remarked that the $b$-trigger is less and less needed as the considered resonance gets heavier. In fact in the following we  consider also masses of the stop that should be well within the range of applicability of standard multi-jet triggers.

The tagging of $b$-jets is of course very important to isolate our signal and in fact is the main handle that we use to improve the bounds from \cite{CMS-PAS-EXO-11-016,CMSCollaboration:2013ui,ATLAS-Collaboration:2012kc,Aad:2011ly} recasted for the RPV stop.
For illustration we take the tagging efficiency to be $\epsilon_{b}=0.66$ and for simplicity we assume it not to depend on the value of $p_{T}$ and $\eta$ of the tagged jet. Since we only consider central jets, in a relatively narrow range of $p_{T}$, these effects should be small. Taking as reference the recent assessment of the $b$-tagging efficiency~\cite{ATLAS-CONF-2012-097,CMS-Collaboration:2012uq} we take the $b$-tagging probability to be less than 1\% for a light jet and 10\% for a charm jet.

\subsection{Analysis}
In this section we discuss the discovery potential of RPV stop pair-production in the process of Eq.~\eqref{process} at the LHC at $\sqrt{s}=8$~TeV. For simplicity we assume that the stop decays into  a $b$ and a $s$ quark 100\% of the times. For our analysis we take two reference $\tilde{t}$ masses $m_{\tilde{t}}=200$~GeV and 100~GeV. For $m_{\tilde{t}}=200$~GeV the analysis can be carried out relying solely on triggers that have already been employed for this kind of searches. For $m_{\tilde{t}}=100$~GeV instead it seems necessary to use $b$-tagging information to reduce the trigger thresholds down to 50~GeV in the offline reconstruction of the fourth jet of the event\footnote{Although we are not aware of any publicly available analysis that uses a similar  trigger on the 2012 data, an item  very close to our trigger should be available in the trigger menu of the ATLAS experiment~\cite{Coccaro}.
}.
 The chosen values of the stop mass are well within the range indicated by the naturalness of the electroweak scale, thus the proposed analysis probes a very interesting scenario for Natural RPV SUSY. 

The main SM backgrounds to our signal are QCD multi-jet production and $t\bar{t}$ production. 
With the aforementioned $b$-jet misidentification rate we checked that the QCD multi-jet production is dominantly due to \mbox{$pp\to b\bar{b} jj$}  and in the following we  consider only this process for the QCD background.
For the $t\bar{t}$ background we found that most of the rate comes from top decays with at least one $W$ decaying hadronically. In this case we  retain all the $t\bar{t}$ decay modes in our analysis. 

To isolate the signal we  exploit its resonant structure as opposed to the incoherent nature of the QCD multi-jet background and the different resonant structure of  $t\bar{t}$ production. In particular we attempt to reconstruct the two stop resonances and we  identify the signal as a bump over a smooth distribution of the reconstructed resonance mass obtained with our reconstruction procedure.
We  put particular care in devising cuts that avoid as much as possible to produce a background distribution  that is too similar in shape to the signal. We work especially to obtain two distributions whose peaks are resolvable with the expected di-jet invariant mass resolution. Furthermore we aim to give a smoothly falling shape to the background in the vicinity of the expected bump, to make it more easily detectable by a suitable bump-hunting procedure (see, for instance, Ref.~\cite{Choudalakis:2011yf}). 

For our analysis we consider only the four hardest jets found in each event within the acceptance selections (defined later). In the case of the signal these jets should correspond to the hadronic activity that arises from the decay of the colored stop resonances. Other jets are not taken into account in the analysis since they are more likely originated from soft processes that accompany the pair production of stops.
Inspired by previous studies  we consider two possible algorithms to pair the four final state jets in order to isolate the signal by fully reconstructing the two intermediate resonances:
\bit
\item {\it Mass pairing} \cite{CMS-PAS-EXO-11-016}: choose the two jet pairs $(ab)$ and $(cd)$ that minimize the quantity
\be\l{eq10}
\delta_{m}=\f{\left|m_{ab}-m_{cd}\right|}{m_{ab}+m_{cd}}\,;
\ee
\item {\it $\Delta R$ pairing} \cite{ATLAS-Collaboration:2012kc,Aad:2011ly}: choose the two jet pairs $(ab)$ and $(cd)$ that minimize the quantity
\be\l{eq11}
\delta_{\Delta R}=\left|\Delta R_{ab}-1\right|+\left|\Delta R_{cd}-1\right|\,,
\ee
\eit
where $m_{ij}$ is the invariant mass of the $(ij)$ pair and $\Delta R_{ij}=\sqrt{\left|\Delta \eta_{ij}\right|^{2}+\left|\Delta \phi_{ij}\right|^{2}}$ is the angular distance between two jets defined as a function of the azimutal angle $\phi$ and the pseudorapidity $\eta=-\log[\tan\f{\theta}{2}]$.

The mass pairing attempts to reconstruct the resonances by insisting on the pair-production of two resonances of identical mass. The $\Delta R$ pairing instead exploits the fact that in the laboratory frame the decay products of each resonance tend to be collimated. This is especially the case for light resonances that are likely to be produced with a large boost. Furthermore the collimation of the decay products is enhanced when considering jets of $p_{T}$ comparable to the mass of the resonance as  we do in the following to isolate our signal from the background.

We  define the following kinematic variables that are used in our analysis:
\bes\l{eq12}
\begin{align}
& \dst m_{\text{best}}=\f{m_{ab}+m_{cd}}{2}\,,\\
& \dst \Delta \eta_{\text{best}}=\f{|\Delta \eta_{ab}|+|\Delta \eta_{cd}|}{2}\,,\\
& \dst \Delta \phi_{\text{best}}=\f{\Delta \phi_{ab}+\Delta \phi_{cd}}{2}\,,\\
& \dst \Delta R_{\text{best}}=\f{\Delta R_{ab}+\Delta R_{cd}}{2}\,,\\
& \dst \cos\theta^{*}=\f{p_{z\,a}^{\text{cm}}+p_{z\,b}^{\text{cm}}}{\left|\mathbf{p}_{a}^{\text{cm}}+\mathbf{p}_{b}^{\text{cm}}\right|}=\f{p_{z\,c}^{\text{cm}}+p_{z\,d}^{\text{cm}}}{\left|\mathbf{p}_{c}^{\text{cm}}+\mathbf{p}_{d}^{\text{cm}}\right|}\,,
\end{align}
\ees
where $\theta^{*}$ is the scattering angle of the reconstructed di-jet resonance in the center of mass frame, computed in the approximation of zero $p_{T}$ of the center of mass. 

We have observed that some crucial quantities such as $m_{\text{best}}$ are sensitive to detector effects and therefore have performed our analysis with the hard cross section computed at Leading Order (LO) with Madgraph5 ~\cite{Alwall:2011fk} and the PDF set CTEQ6L1~\cite{Pumplin:2002vw}, soft QCD radiation simulated with Pythia~8~\cite{Sjostrand:2007bk} and detector response parametrized with  Delphes~2.0~\cite{Ovyn:2009ys}. 
Jets have been made with Fastjet~2~\cite{Cacciari:2011rt,Cacciari:2006vn} using the anti-$k_{T}$ algorithm with parameter $R=0.6$ \cite{Cacciari:2008hb}.

In order to verify that our simulation is consistent with experimental results we reproduced the ATLAS search for pair produced colored resonances in the four jet final state of Ref.~\cite{Aad:2011ly}. The ATLAS results of Ref.~\cite{Aad:2011ly} were obtained using ATLAS full detector simulation on the particle level prediction from a MLM~\cite{Alwall:2007kz} matched computation. After all kinematic selections we agree with the simulation used in the experimental analysis at the 30\% level.  Since the final state we are studying in the present work only differs from the one of the aforementioned ATLAS analysis by the relevance of the \mbox{$b$-tagging}, which we take into account with a flat efficiency factor, we expect our predictions  to be reasonably realistic.

For each of the two cases, {\it Mass pairing} and $\Delta R$ {\it pairing}, we devised an optimized strategy to isolate the signal. In doing so we explored selection strategies involving, among others\footnote{In particular we considered extra selection involving the transverse sphericity, the sphericity and the quantity $\Delta$ defined in Ref.~\cite{CMS-PAS-EXO-11-016}.}, the quantities defined in Eqs.~\eqref{eq12}. For the $\Delta R$ {\it pairing} we are more easily able to obtain a smoothly falling shape for the background in the vicinity of the sought stop mass. Moreover, with the $\Delta R$ {\it pairing} we find a better signal over background ratio both in the overall rates and for $m_{\text{best}}$ in the vicinity of the relevant stop mass. Therefore the $\Delta R$ {\it pairing} seems to perform overall better than the {\it Mass pairing} in terms of the sharpness of the isolation of the new physics contribution. This gain in clarity of the observation may not always correspond to a better statistical significance, depending on the total efficiency of the selections. However we decided to insist on the $\Delta R$ {\it pairing} to pursue as much as possible the clarity of the observation of new physics. In presence of the large uncertainties that are inherent in the QCD background this seems to be the most robust way to assess the presence of new physics.


\begin{figure*}
\hspace{-2.4mm}\includegraphics[width=0.99 \columnwidth]{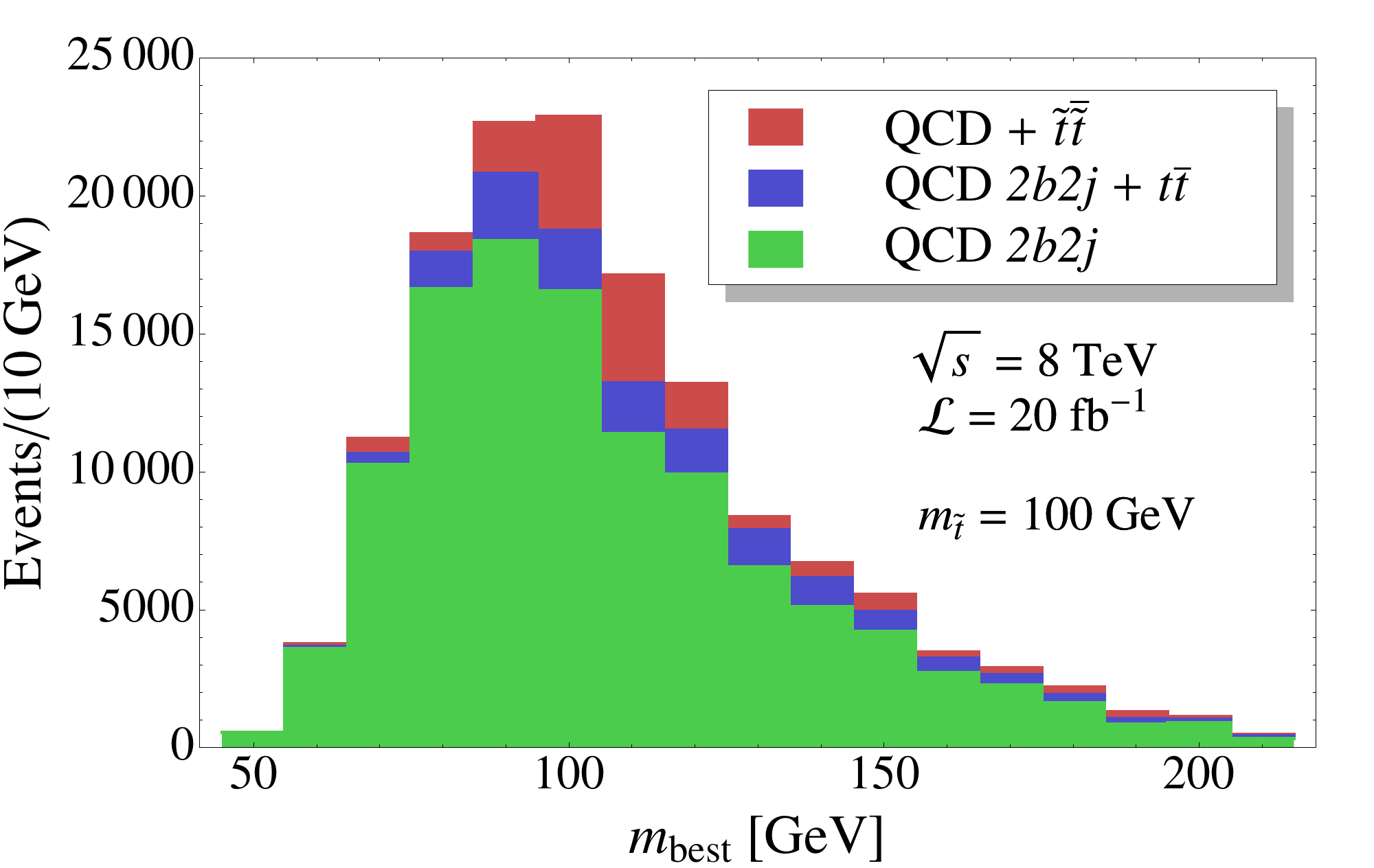}\includegraphics[width=0.95 \columnwidth]{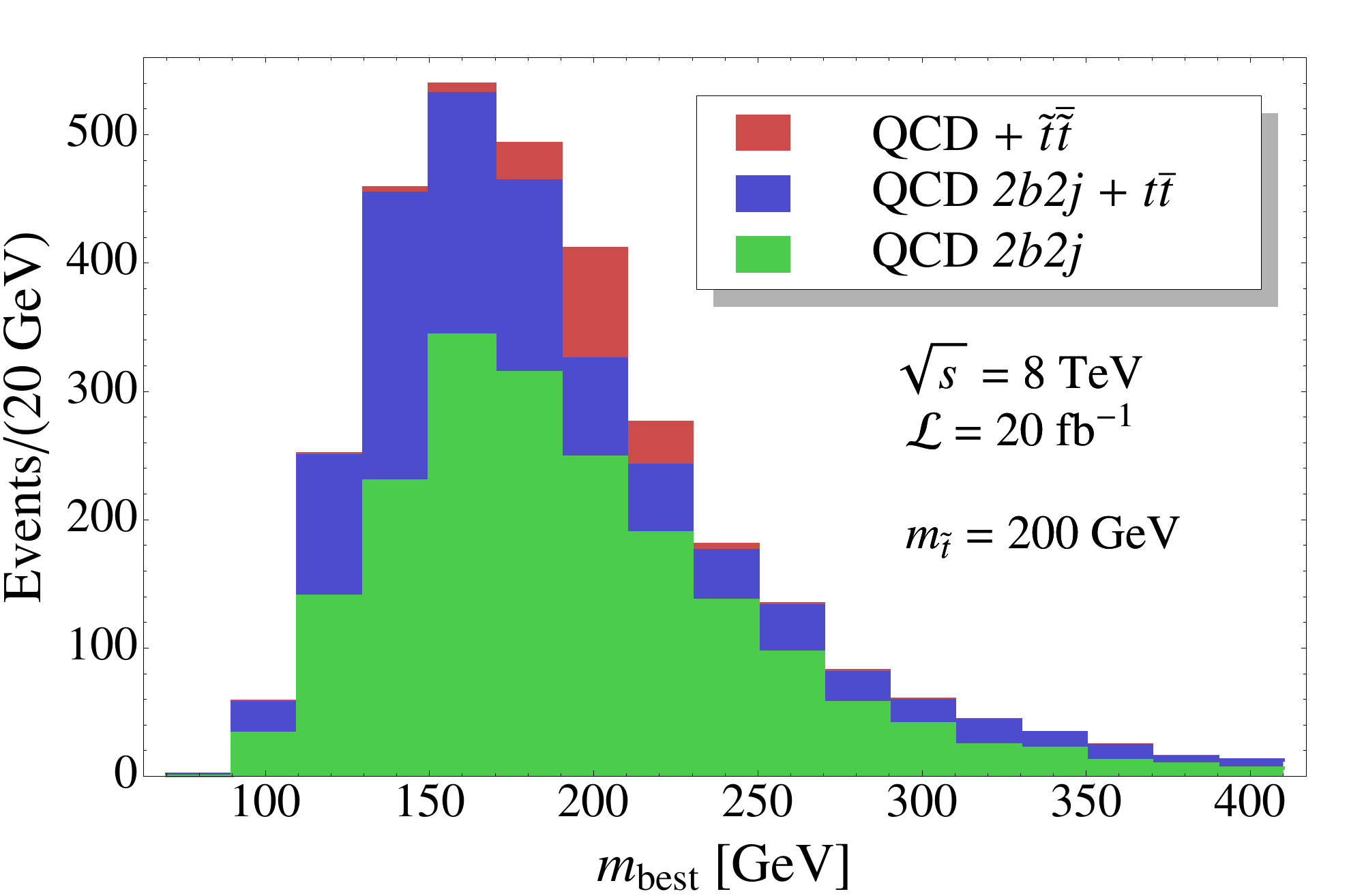}\\
\hspace{0mm}\includegraphics[width=0.97 \columnwidth]{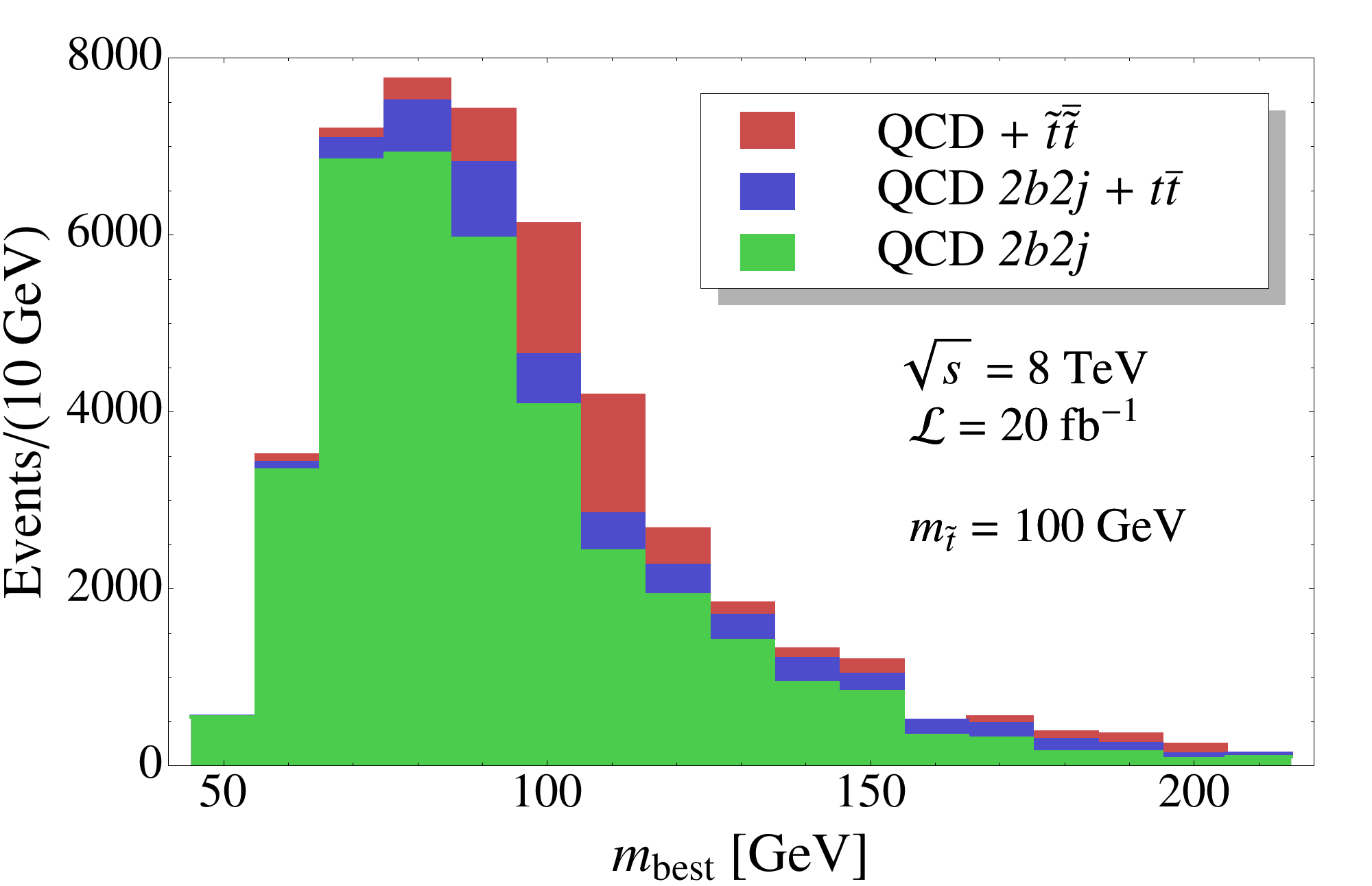}\includegraphics[width=0.95 \columnwidth]{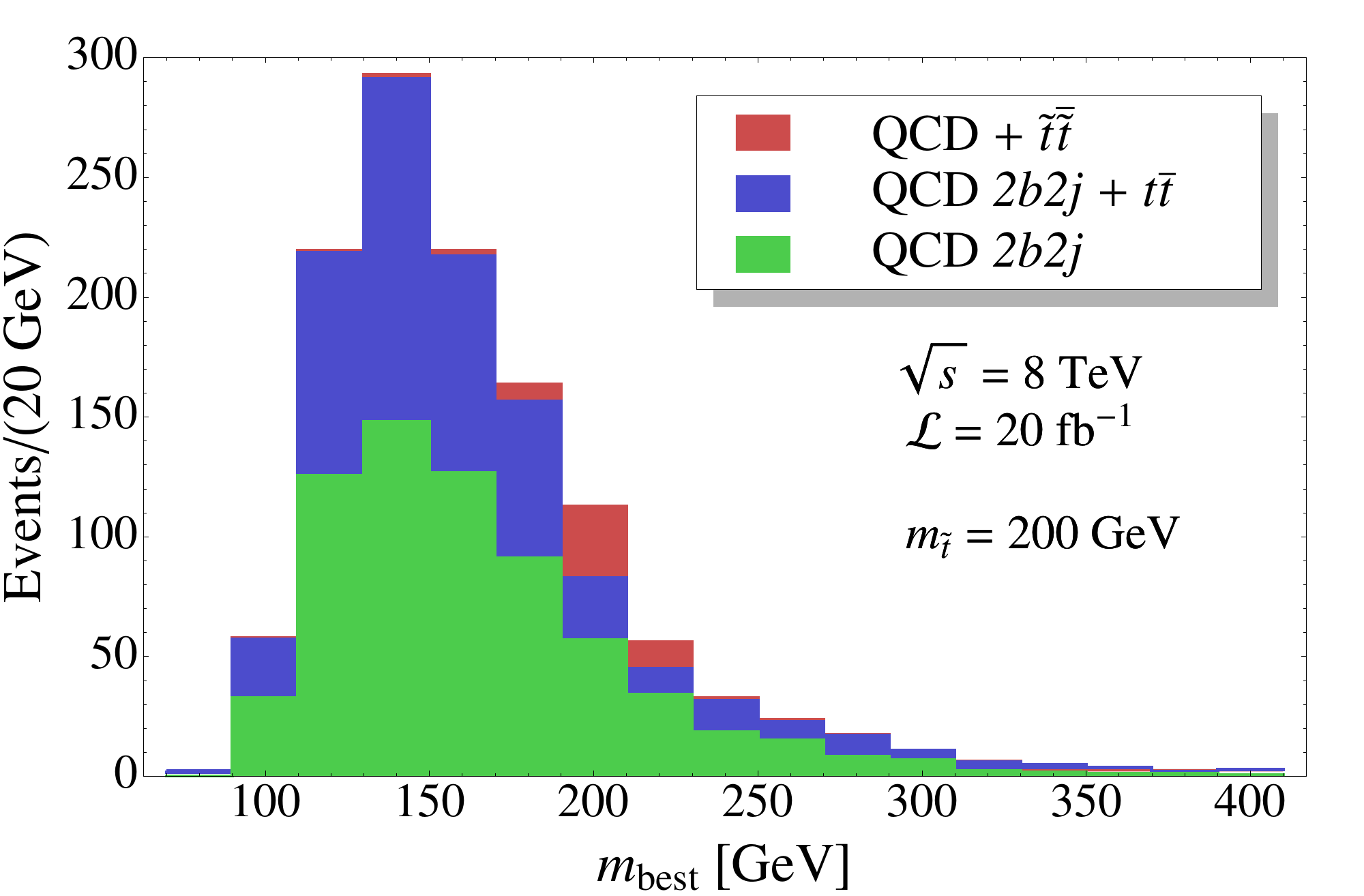}
\caption{\label{SandB} $m_{\text{best}}$ distribution for the two analyses for $m_{\tilde{t}}=100\gev$ (left) and  $m_{\tilde{t}}=200\gev$ (right). The upper row corresponds to a loose selection $\Delta R_{\textrm{best}}<1.5$  that can give higher significance at the price of less resolved shapes for the background and the signal. The lower row corresponds to a tighter selection $\Delta R_{\textrm{best}}<1$ that privileges a sharper separation of signal and background.}
\end{figure*}


In our analysis  we consider events with at least four jets satisfying the following requirements:

\vspace{2mm}
\begin{empheq}[box=\fbox]{equation}
\bry{lll}
p_{T,j}>{m_{\tilde{t}} \over 2}\,\,,& |\eta_{j}|<2.8\,, &\Delta R_{jj}>0.7\,,\vspace{2mm}\\
\delta_{m}<0.075\,, & \left|\cos\theta^{*}\right|<0.4\,, &\Delta R_{\text{best}}<1.5\,,\vspace{2mm}\\
\Delta\eta_{\text{best}}<0.8\,. & &
\ery\l{eq16}
\end{empheq}

The signal is identified as a bump in the $m_{\text{best}}$ distribution as shown in Figure \ref{SandB} (upper row). 
The bin size chosen to plot the $m_{\text{best}}$ distribution corresponds to $10\%$ of the resonance mass that we consider. This choice is consistent with the expected experimental resolution on the di-jet invariant mass.
The efficiencies of the selections and the LO cross sections used in the calculation are collected in Table \ref{Table1} for the two choices $m_{\tilde{t}}=100\gev$ and $m_{\tilde{t}}=200\gev$.
It is clear from the Figure that there can be several bins with a signal over background ratio {30\%}. 
Given the sensitivity to multi jet signals that the ATLAS and CMS experiments have demonstrated~\cite{ATLAS-Collaboration:2012kc,CMSCollaboration:2013ui} we expect this signal to be observable.
We estimate that with a luminosity of about 20/fb the observation can be made at about the 5$\sigma$ confidence level for both choices of the stop mass that we consider. 

However we notice that the separation of the background and signal peaks is marginal with the expected resolution in the di-jet invariant mass. Depending on the performances of the experiments on this quantity it may be desirable to obtain a better separation between the signal and the background peaks. To achieve this we can enforce a tighter cut on $\Delta R_{\text{best}}$, which pushes the background peak towards smaller values of $m_{\text{best}}$. To display the attainable peak separation we show, in the lower row of Figure \ref{SandB}, the results for 
$$\Delta R_{\text{best}}<1\,.$$
The signal in this case can be clearly identified as a distinct peak on a smoothly falling background shape, a situation that improves the possibility to observe or exclude the new physics. We notice however that tightening the requirement on $\Delta R_{\text{best}}$ in general reduces the achievable statistical significance of the observation
~\footnote{NLO corrections to the production rates are in general not small and a modest increase in the significance is expected to arise from these corrections.}. We leave the detailed study of the best balance between the significance and the reduction of uncertainties in the observation to the experimental collaborations. Most of the result of this kind of optimization depends in fact on the actual detailed performances of the detectors.
\begin{table*}[!th]
\begin{minipage}{\textwidth}
\begin{center}
\caption{\label{Table1} Signal and background efficiencies after the kinematic selections of Eq.~\eqref{eq16} at the LHC at $\sqrt{s}=8$ TeV using the {\it $\Delta R$ pairing}. $\epsilon^{(1)}$ is the efficiency of the single cut independently of the others, $\epsilon$ is the global efficiency of the cuts at and above that line in the table, $\epsilon_{i\to i+1}$ is the efficiency of the cut with respect to the previous line. The numbers in parenthesis are the total cross sections computed at LO with Madgraph5. Corrections to the overall rate from the NLO in QCD have been computed and give $K_{\rm NLO}\simeq 1.5$ both for the signal (see for instance Ref.~\cite{Beenakker:2011wz}) and for multi-jet backgrounds (see for instance Refs.~\cite{Greiner:2011cn,Bern:2011vg}), therefore they are not expected to alter our results.}
\begin{tabular}{c|ccc|ccc|ccc}
  &\multicolumn{9}{c}{$m_{\tilde{t}}=100\gev$ - \it $\Delta R$ pairing}\\
 \hline
&\multicolumn{3}{c}{$\tilde{t}\ovl{\tilde{t}}$ (314 pb)}
&\multicolumn{3}{c}{QCD $b\bar{b}jj$ (8826 pb)\footnote{This QCD cross section is computed taking $p_{T}>35$ GeV, $|\eta|<3.5$ and $\Delta R > 0.4$ for the matrix element computation. 
}}
&\multicolumn{3}{c}{$t\bar{t}$(135 pb)}\\
\hline
Selection   & $\epsilon^{(1)}$ & $\epsilon$ & $\epsilon_{i\to i+1}$   & $\epsilon^{(1)}$ & $\epsilon$ & $\epsilon_{i\to i+1}$ & $\epsilon^{(1)}$ & $\epsilon$ & $\epsilon_{i\to i+1}$\\
\hline
$\eta<2.8$ 				& 0.81 & 0.81 & - 		& 0.82 & 0.82 & -  		& 0.88 & 0.88 & -  \\
$p_{T}>50$ GeV 			& 0.16 & 0.16 & 0.19  	& 0.15 & 0.15 & 0.18 		& 0.38 & 0.38 & 0.43 \\
$\Delta R>0.7$ 			        	& 0.78 & 0.15 & 0.95		& 0.79 & 0.14 & 0.96		& 0.85 & 0.36 & 0.94 \\
$b$-tags $= 2$				& 0.44 & 0.064 & 0.44 	& 0.44 & 0.062 & 0.44	& 0.44 & 0.15 & 0.44 \\
$\delta_{m}<0.075$	 		& 0.13 & 0.010 & 0.16	& 0.11 & 0.0085 & 0.14	& 0.15 & 0.026 & 0.17 \\
$\left|\cos\theta^{*}\right|<0.4$	& 0.33 & 0.0047 & 0.46	& 0.19 & 0.0021 & 0.24	& 0.36 & 0.026 & 0.45 \\
$\Delta\eta_{\text{best}}<0.8$    	& 0.31 & 0.0030 & 0.64	& 0.23 & 0.00077 & 0.38	& 0.37 & 0.0069 & 0.60 \\
$\Delta R_{\text{best}}<1.5$	& 0.25 & 0.0025  & 0.85	& 0.19 & 0.00063 & 0.82	& 0.31 & 0.0056 & 0.81  \\
$\Delta R_{\text{best}}<1$	         & 0.031 & 0.00080  & 0.32 	& 0.030 & 0.00020 & 0.32 	& 0.043 & 0.0016 & 0.28\\
\end{tabular}\vspace{4mm}
\begin{tabular}{c|ccc|ccc|ccc}
&\multicolumn{9}{c}{$m_{\tilde{t}}=200\gev$ - \it $\Delta R$ pairing}\\
 \hline
&\multicolumn{3}{c}{$\tilde{t}\ovl{\tilde{t}}$ (9.1 pb)}
&\multicolumn{3}{c}{QCD $b\bar{b}jj$ (136 pb)\footnote{This QCD cross section is computed taking $p_{T}>75$~GeV, $|\eta|<3.5$ and $\Delta R > 0.4$ for the matrix element computation. 
}}
&\multicolumn{3}{c}{$t\bar{t}$(135 pb)}\\
\hline
Selection  & $\epsilon^{(1)}$ & $\epsilon$ & $\epsilon_{i\to i+1}$   & $\epsilon^{(1)}$ & $\epsilon$ & $\epsilon_{i\to i+1}$ & $\epsilon^{(1)}$ & $\epsilon$ & $\epsilon_{i\to i+1}$\\
\hline
$\eta<2.8$ 				& 0.16 & 0.16 & - 		& 0.94 & 0.036 & - 		& 0.88 & 0.88 & -  \\
$p_{T}>100$ GeV 			& 0.026 & 0.026 & 0.16 	& 0.13 & 0.13 & 0.14 		& 0.0031 & 0.31 & 0.035  \\
$\Delta R>0.7$ 				& 0.15 & 0.035 & 0.95 	& 0.88 & 0.12 & 0.93  	& 0.85 & 0.027 & 0.87\\
$b$-tags $= 2$				& 0.44 & 0.011 & 0.44 	& 0.44 & 0.52 & 0.44 		& 0.44 & 0.012 & 0.44 \\
$\delta_{m}<0.075$	 		& 0.036 & 0.0031 & 0.29	& 0.12 & 0.0072 & 0.14	& 0.15 & 0.0015 & 0.13 \\
$\left|\cos\theta^{*}\right|<0.4$	& 0.096 & 0.0018 & 0.57	& 0.25 & 0.0021 & 0.29 	& 0.36 & 0.00066 & 0.45\\
$\Delta\eta_{\text{best}}<0.8$    	& 0.078 & 0.0013 & 0.73	& 0.29 & 0.00084 & 0.41	& 0.38 & 0.00044 & 0.66\\
$\Delta R_{\text{best}}<1.5$	& 0.075 & 0.0011  & 0.85 	& 0.26 & 0.00071 & 0.84 	& 0.31 & 0.00038 & 0.86  \\
$\Delta R_{\text{best}}<1$		& 0.012 & 0.00031  & 0.29	& 0.046 & 0.00025 & 0.35 	& 0.043 & 0.00019 & 0.49  \\
\end{tabular}
\end{center}
\end{minipage}
\end{table*}
\section{Conclusions}\label{conclusions}
In this paper we discussed the discovery potential of the LHC of pair-production of heavy flavored di-jet resonances.
We studied the well motivated example of the RPV stop in detail. However our analysis, with minor modifications, can be applied to other types of new physics that result in the $2b2j$ final state. Our study was carried out including the effect of QCD radiation and detector reconstruction effects on the final state of the hard collision. We employed the angular method of Refs.~\cite{ATLAS-Collaboration:2012kc,Aad:2011ly} to identify the candidate reconstructed stop resonances and showed that, with suitable selections, the production of stops can result in a bump in the distribution of  $m_{\textrm{best}}$, the mass of the candidate resonances. We also showed that the shape of the $m_{\text{best}}$ background distribution can be modified to be more favorable for the identification of the signal by applying cuts on suitable angular variables.

In particular we identified the kinematic variable $\Delta R_{\text{best}}$ as the quantity to better control the background shape and enforce a  separation between the signal and the background peaks sufficient to be resolved with the experimental resolution on the invariant mass of the system of two jets. Figure \ref{SandB} shows the difference in the result that can be induced by adjusting the cut on $\Delta R_{\text{best}}$. The same Figure also clearly shows that the signal of stop production can be observed with a signal over background ratio of the order of $10\%$ in several bins around the mass of the resonance. 

For the case of a lighter stop, for instance the case of $m_{\tilde{t}}=100\gev$ that we studied explicitly, we highlighted the importance of having low thresholds in the triggers in order to retain most of the signal from light stop production. In this respect we welcome the advent of $b$-jet identification at trigger level in the LHC experiments that allows us to keep low trigger thresholds even in a high instaneous luminosity environment such as the 2012 run of the LHC. Alternatively, for light stop masses the events could pass a trigger for multiple jets thanks to the presence of hard, hence costly in terms of production rate, extra QCD radiation. While this is certainly an interesting way to look for light colored resonances at the LHC we did not study this possibility which is, however, an interesting check for observations carried out with our method.
Finally we encourage the experiments to broaden their program for the search of new physics connected with the naturalness of the electroweak scale. In particular we encourage them to carry out  heavy flavored multi-jet searches that can probe large parts of the parameters space of scenarios of natural supersymmetric theories with R-parity violation. 

\begin{acknowledgements}
We thank Dinko Ferencek, Shahram Rahatlou, Kai Yi for clarifications on the multi-jet searches of CMS and Andrea Coccaro for discussions on the current and future trigger in ATLAS. We also thank Roberto Contino,  Andrey Katz and Daniel Stolarski for discussions. We thank the CERN Theory Division for hospitality and support while this research was carried out. 
RF thanks the Galileo Galilei Institute for hospitality and support during the completion of this work. 
The work of RF is supported by the NSF Grants PHY-0910467 and PHY-0968854 and by the Maryland Center for Fundamental Physics.
The work of RT was partly supported by the Spanish MICINN  under grants CPAN CSD2007-00042 (Consolider-Ingenio 2010 Programme) and FPA2010-17747, by the Community of Madrid under grant HEPHACOS S2009/ESP-1473, by the Research Executive Agency (REA) of the European Union under the Grant Agreement number PITN-GA-2010-264564 (LHCPhenoNet) and by the ERC Advanced Grant no. 267985, Electroweak Symmetry Breaking, Flavour and Dark Matter: One Solution for Three Mysteries (DaMeSyFla).
\end{acknowledgements}

\bibliographystyle{mine}
\bibliography{bibliography}

\end{document}